\documentclass[12pt,preprint]{aastex}
\usepackage[T1]{fontenc} 
\usepackage[english]{babel}    % langue 
\RequirePackage{xspace}
\RequirePackage{amsmath}
\RequirePackage{amsfonts}
\RequirePackage{amssymb}
\usepackage{floatflt}
\RequirePackage{amsfonts}
\RequirePackage{amssymb}

\def\ifundefined#1{\expandafter\ifx\csname#1\endcsname\relax}

\def\la{\mathrel{\hbox{\rlap{\hbox{\lower4pt\hbox{$\sim$}}}\hbox{$<$}}}}
\def\ga{\mathrel{\hbox{\rlap{\hbox{\lower4pt\hbox{$\sim$}}}\hbox{$>$}}}}

\newcommand{\be}{\begin{eqnarray}}
\newcommand{\ee}{\end{eqnarray}}

\ifundefined{ensuremath}\def\ensuremath#1{\relax\ifmmode{#1}}
\else${#1}$\fi\else\relax\fi
\ifundefined{nuc}\def\nuc#1#2{\relax\ifmmode{}^{#1}{\protect\textrm{#2}}
\else${}^{#1}$#2\fi}\else\relax\fi
\newcommand{\RSi}{$\Re_{Si}$\xspace}
\newcommand{\RCa}{$\Re_{Ca}$\xspace}

\newcommand{\scname}[1]{\textsc{#1}}

\newcommand{\SiIIred}{\SiII$\lambda 6355$\xspace}
\newcommand{\SiIIblue}{\SiII$\lambda 5972$\xspace}
\newcommand{\SiII}{Si~\scname{ii}\xspace}
\newcommand{\CaII}{Ca~\scname{ii}\xspace}

\begin{document}

\title{Quantitative Spectroscopy of Supernovae for Dark Energy Studies}

\author{ 
  E.~Baron,\altaffilmark{1,2} %\email{baron@nhn.ou.edu}
David
Branch,\altaffilmark{1} %\email{branch@nhn.ou.edu}
David Jeffery,\altaffilmark{1} %\email{jeffery@nhn.ou.edu}
  Peter Nugent,\altaffilmark{2} %\email{penugent@lbl.gov}
  Rollin Thomas,\altaffilmark{3} %\email{rcthomas@lbl.gov}
  Sebastien Bongard,\altaffilmark{3,4} %\email{bongard@in2p3.fr},
Peter H.~Hauschildt,\altaffilmark{5} %\email{yeti@hs.uni-hamburg.de}
Daniel Kasen,\altaffilmark{6} %\email{kasen@pha.jhu.edu} 
and Dimitri Mihalas\altaffilmark{7}%\email{dmihalas@lanl.gov}
 }

\altaffiltext{1}{Homer L.~Dodge Department of Physics and Astronomy,
  University of 
Oklahoma, 440 West Brooks, Rm.~100, Norman, OK 73019-2061, USA}

\altaffiltext{2}{Computational Research Division, Lawrence Berkeley
  National Laboratory, MS 50F-1650, 1 Cyclotron Rd, Berkeley, CA
  94720-8139 USA}

\altaffiltext{3}{Lawrence Berkeley
  National Laboratory, MS 50R-5008, 1 Cyclotron Rd, Berkeley, CA
  94720 USA}

\altaffiltext{4}{Institute de Physique Nucl\'eaire Lyon, B\^atiment Paul Dirac
Universit\'e Claude Bernard Lyon-1
Domaine scientifique de la Doua
4, rue Enrico Fermi
69622 Villeurbanne cedex, France}

\altaffiltext{5}{Hamburger Sternwarte, Gojenbergsweg 112,
21029 Hamburg, Germany}

\altaffiltext{6}{Department of Physics and Astronomy, Johns Hopkins University,
Baltimore MD, 21218}

\altaffiltext{7}{Applied Physics Division, X-3, MS-F644, Los Alamos
  National Laboratory, Los Alamos, NM 87545 USA}

\vfill\eject

\section{Project Summary}

Detailed quantitative spectroscopy of Type Ia supernovae (SNe~Ia)
provides crucial information needed to minimize systematic effects in
both ongoing SNe~Ia observational programs such as the Nearby
Supernova Factory, ESSENCE, and the SuperNova Legacy
Survey  (SNLS) and in proposed JDEM missions such as SNAP, JEDI, and DESTINY.

Quantitative spectroscopy is mandatory to quantify and understand the
observational strategy of comparing ``like versus like''. It allows us
to explore evolutionary effects, from variations in progenitor
metallicity to variations in progenitor age, to variations in dust
with cosmological epoch. It also allows us to interpret and quantify
the effects of asphericity, as well as different amounts of mixing in the
thermonuclear explosion. 

While all proposed cosmological measurements will be based on empirical
calibrations, these calibrations must be interpreted and evaluated in
terms of theoretical explosion models. Here quantitative spectroscopy
is required, since explosion models can only be tested in
detail by  direct comparison of detailed NLTE synthetic spectra with
observed spectra. 

Additionally, SNe IIP can be used as complementary cosmological probes
via the spectral fitting expanding atmosphere method (SEAM) that we
have developed. The SEAM method in principle can be used for distance
determinations to much higher $z$ than Type Ia supernovae.

We intend to model in detail the current, rapidly growing, database
of SNe Ia and SNe IIP. Much of the data is immediately available in
our public spectral and photometric database SUSPECT, which is widely used
throughout the astronomical community.

We bring to this effort a variety of complementary synthetic spectra
modeling capabilities: the fast parameterized 1-D code SYNOW; BRUTE, a
3-D Monte-Carlo with similar assumptions to SYNOW; a 3-D Monte-Carlo
spectropolarimetry code, SYNPOL; and the generalized full NLTE, fully
relativistic stellar atmosphere code PHOENIX (which is being
generalized to 3-D).

\section{Cosmology from Supernovae}

While indirect evidence for the cosmological acceleration can be
deduced from a combination of studies of the cosmic microwave
background and large scale structure
\citep{efstat02,map03,eisensteinbo05}, distance measurements to 
supernovae provide a valuable direct and model independent tracer of
the evolution of the expansion scale factor necessary to constrain the
nature of the proposed dark energy.  The mystery of dark energy lies
at the crossroads of astronomy and fundamental physics: the former is
tasked with measuring its properties and the latter with explaining
its origin.

Presently, supernova measurements of the cosmological parameters are
no longer limited by statistical uncertainties, but systematic
uncertainties are the dominant source of error \citep[see][for a
recent analysis]{knopetal03}.  These include the effects of evolution (do
SNe~Ia behave in the same way in the early universe?), the effect of
intergalactic dust on the apparent brightness of the SNe~Ia, and
knowledge of the spectral energy distribution as a function of light
curve phase (especially in the UV where are current data sets are
quite limited).

Recently major ground-based observational programs have begun: the
Nearby SuperNova Factory \citep[see][]{aldering_nearby,
  nugent_nearby}, the European Supernova Research Training Network
(RTN), the Carnegie Supernova Project (CSP), ESSENCE, and the
SuperNova Legacy Survey. Their goals are to improve our understanding
of the utility of Type Ia supernovae for cosmological measurements by
refining the nearby Hubble diagram, and to make the first definitive
measurement of the equation of state of the universe using $z < 1$
supernovae.  Many new programs have recently been
undertaken to probe the rest-frame UV region at moderate $z$,
providing sensitivity to metallicity and untested progenitor
physics. SNLS has found striking diversity in the UV behavior that is not
correlated with the normal light curve stretch parameter. As precise
knowledge of the $K$-correction is needed to use SNe~Ia to trace the
deceleration expected beyond $z=$1 \citep{riessetal04a}, understanding
the nature of this diversity is crucial in the quest for measuring
dark energy.  We plan to undertake an extensive theoretical program,
which leverages our participation with both SNLS and the Supernova
Factory, in order to refine our physical understanding of supernovae
(both Type Ia and II) and the potential systematics involved in their
use as cosmological probes for the Joint Dark Energy Mission (JDEM).

In addition to SNe~Ia, the Nearby Supernova Factory will
observe scores of Type IIP supernovae in the Hubble
Flow. These supernovae will provide us with a perfect laboratory to
probe the mechanisms behind these core-collapse events, the energetics
of the explosion, asymmetries in the explosion event and thereby
provide us with an independent tool for precision measurements of the
cosmological parameters.

\begin{figure}
\includegraphics[width=0.7\hsize,angle=90]{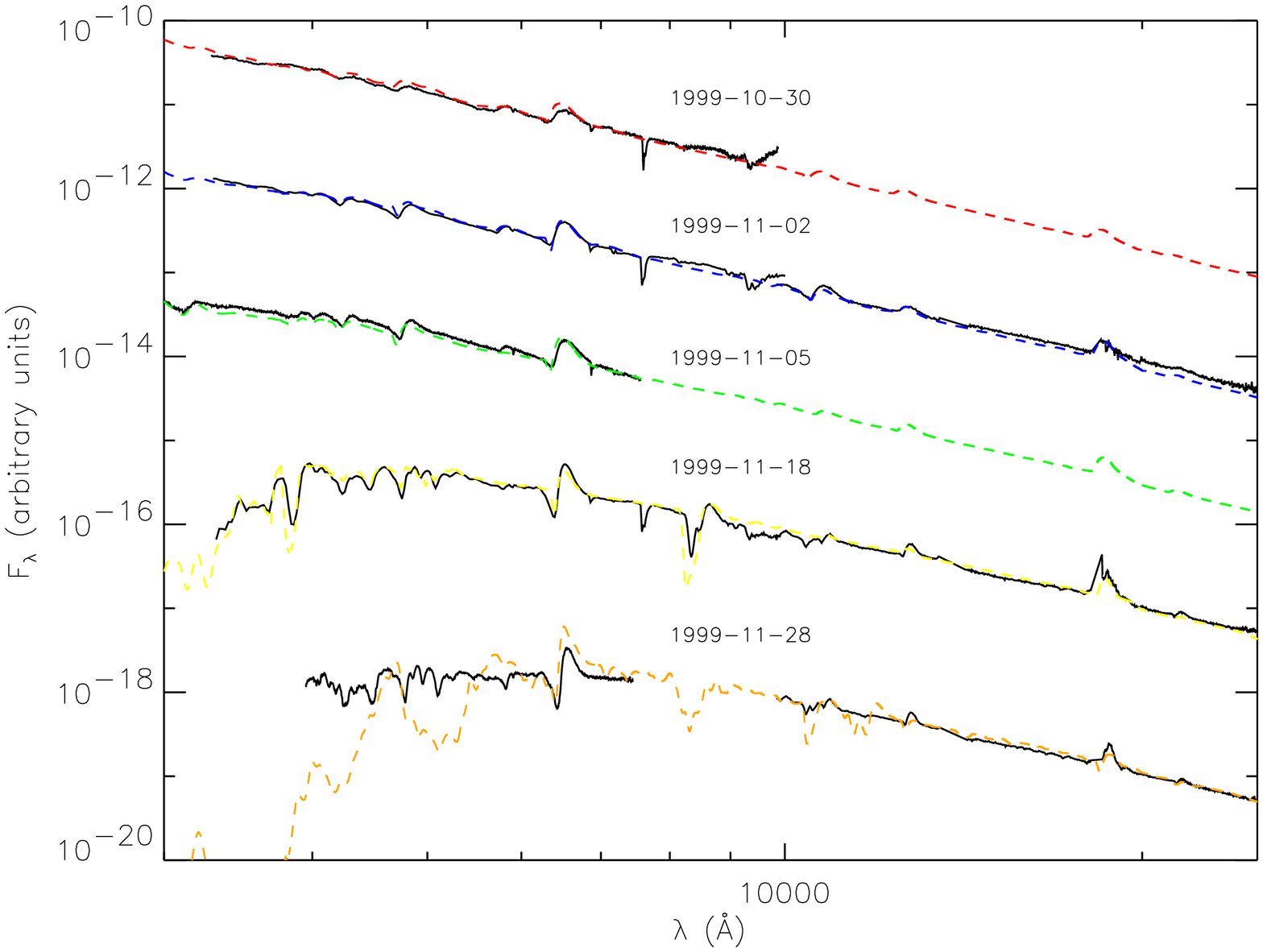}
\caption{\label{fig:fits}The synthetic spectra of SN~1999em (dashed
  lines) are compared to 
  observed spectra (solid lines) at 5 different epochs. The ability to
accurately model the SED allows us to obtain accurate distances via the SEAM
method \citep{bsn99em04}.}
\end{figure}

The SEAM method has shown that accurate distances may be obtained to
SNe~IIP, even when the classical expanding photosphere method fails
\citep[see Fig.~\ref{fig:fits} and][]{bsn99em04}. Another part of the
SN~IIP study is based a correlation between the absolute brightness of
SNe~IIP and the expansion velocities derived from the Fe~II 5169 \AA\
P-Cygni feature observed during their plateau phases
\citep{hp02}. We have refined this method in two ways (P. Nugent
{\it et al.}, 2005, in preparation) and have applied it to five
SNe~IIP at $z < 0.35$.  Improving the accuracy of measuring distances
to SNe~IIP has potential benefits well beyond a systematically
independent measurement of the cosmological parameters based on SNe~Ia
or other methods. Several plausible models for the time evolution of
the dark energy require distance measures to $z \simeq 2$ and beyond. At
such large redshifts both weak lensing and SNe\,Ia may become
ineffective probes, the latter due to the drop-off in rates suggested
by recent work \citep{strolger04}. Current models
for the cosmic star-formation history predict an abundant source of
core-collapse at these epochs and future facilities, such as JDEM, in
concert with the James Webb Space Telescope (JWST) or the Thirty Meter
Telescope, could potentially use SNe~IIP to determine distances at
these epochs.

\emph{Spectrum synthesis computations provide the only
way to study this wealth of data and use it to quantify and correct
for potential systematics and improve the distances measurements to
both SNe~Ia and SNe~IIP.}

\section{Understanding the 3-D Nature of Supernovae}

\begin{figure}
  \includegraphics[width = 0.8\textwidth, clip]{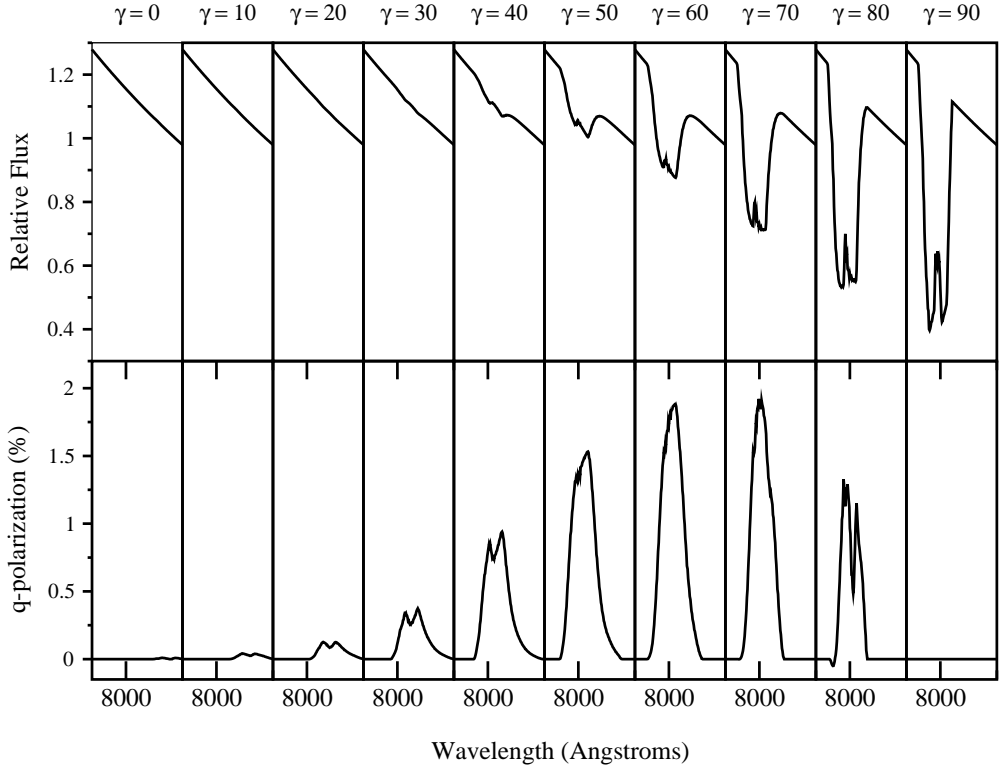}
  \caption[SN01el]{Results of 3D spectrum synthesis using SYNPOL for
    the SN~Ia 2001el. Shown are the fits to the high-velocity Ca~II IR
    triplet for various lines of sight assuming a clump of calcium
    obscures an elliptical photosphere
    \citep[see][]{kasen01el03,KP05}. Given enough observations at
    early times that show high-velocity calcium, we can test whether
    this model is correct and this leads to clues to both the
    progenitor system, explosion models, and potential systematics at
    high redshift.}
  \label{fig:sn01elclump}
\end{figure}

While most SNe~Ia do not show signs of polarization, a subset of them
do. These supernovae will play a role in determining the underlying
progenitor systems/explosion mechanisms for SNe~Ia which is key to
ascertaining potential evolutionary effects with redshift.  Flux and
polarization measurements of SN~2001el \citep{wangetal01el03} clearly
showed polarization across the high-velocity Ca~II IR triplet.  A 3-D
spectopolometric model fit for this object assumes that there is a
blob of calcium at high-velocity over an ellipsoidal atmosphere with
an asphericity of $\approx$ 15\% \citep[see Fig~\ref{fig:sn01elclump}
and][]{kasen01el03}. \citet{KP05} have shown that a gravitationally
confined thermonuclear supernova model can also explain this
polarization signature.  If this is in fact the correct hydrodynamical
explosion model for SNe~Ia, then the parameter space for potential
systematics becomes significantly smaller in their use as standard
candles. Currently there are a wide variety of possible mechanisms to
make a SN~Ia each with its own set of potential evolutionary
systematics. \citet{thomas00cx04} showed that the observed spectral
homogeneity implies that arbitrary asymmetries in SNe~Ia are ruled
out. The only way to test detailed hydrodynamical models of the
explosion event is to confront observations such as those that will be
obtained via the Nearby Supernova Factory with the models via spectrum
synthesis.\emph{The importance of studying these events in 3-D is
  clear from the observations, and therefore every effort must be made
  to achieve this goal.}

\section{\tt PHOENIX}

\label{phoenix}

In order to model astrophysical plasmas under a variety of conditions,
including differential expansion at relativistic velocities found in
supernovae, we have developed a powerful set of working computational
tools which includes the fully relativistic, non-local thermodynamic
equilibrium (NLTE) general stellar atmosphere and spectral synthesis
code {\tt PHOENIX}
\citep{hbmathgesel04,hbjcam99,hbapara97,phhnovetal97,ahscsarev97}. {\tt
  PHOENIX} is a state-of-the-art model atmosphere spectrum synthesis
code which has been developed and maintained by some of us to tackle
science problems ranging from the atmospheres of brown dwarfs, cool
stars, novae and supernovae to active galactic nuclei and extra-solar
planets.  We solve the fully relativistic radiative transport equation
for a variety of spatial boundary conditions in both spherical and
plane-parallel geometries for both continuum and line radiation
simultaneously and self-consistently. We also solve the full
multi-level NLTE transfer and rate equations for a large number of
atomic species, including non-thermal processes. 

\begin{figure}
  \includegraphics[width = 0.8\textwidth, clip]{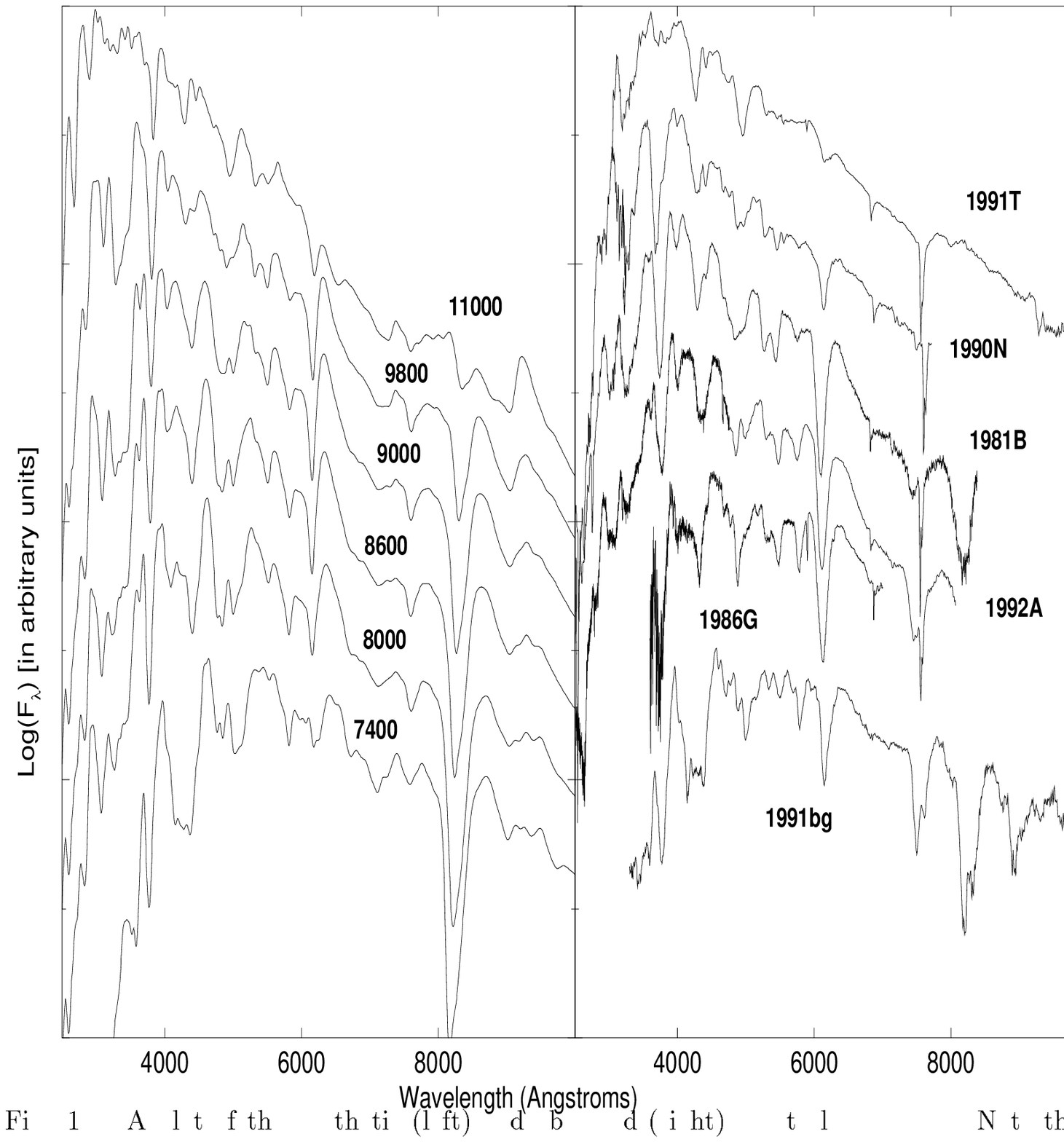}
  \caption[Nug_Seq]{The synthetic spectral sequence of SNe~Ia compared
    with the observed sequence \citep{nugseq95}\label{fig:nugseq}. The
  existence of this spectral sequence and its correlation to the
  photometric one is an ideal example of what can be learned via
  quantitative spectroscopy that is much deeper than ``this
  hydrodynamical model doesn't fit''.}
\end{figure}

\begin{figure}
  \includegraphics[width = 0.8\textwidth, clip]{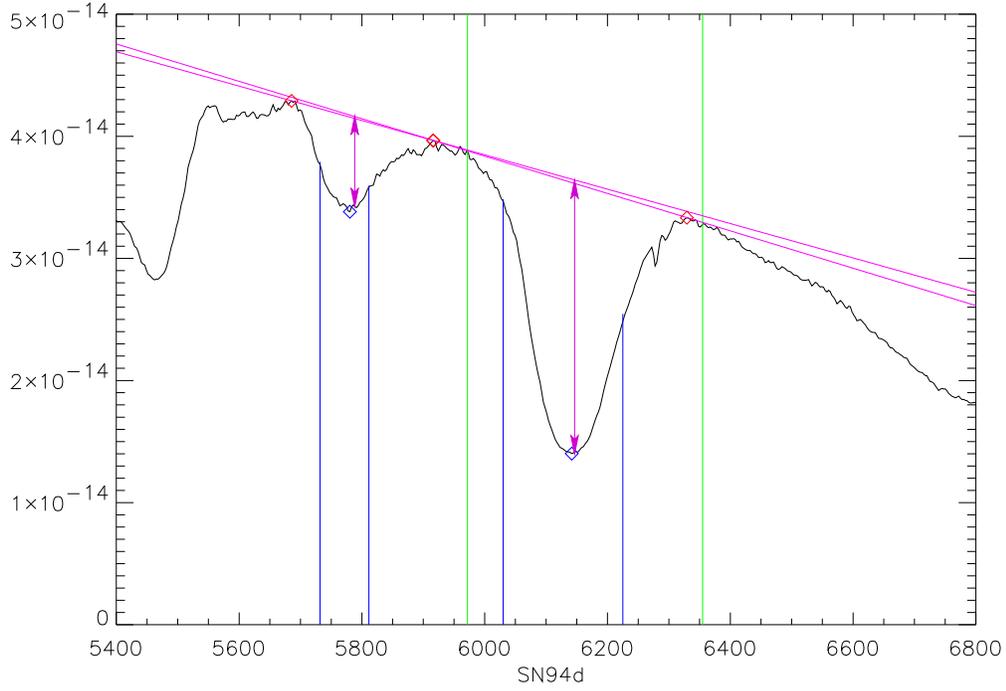}
  \caption[\RSi]{The automated construction used for defining \RSi. Green lines
    locate the \SiIIred and \SiIIblue \SiII lines. The blue lines
    locate the regions where the minima are searched for. The red
    diamonds locate the three maxima used to draw 
    the two purple reference lines, and the two purple arrows
    represent $d_\textrm{red}$ and $d_\textrm{blue}$ used to compute
 $\textrm{\RSi}=\frac{d_\textrm{blue}}{d_\textrm{red}}$.}
  \label{fig:RSiDef}
\end{figure}

\begin{figure}
  \includegraphics[width = 0.8\textwidth, clip]{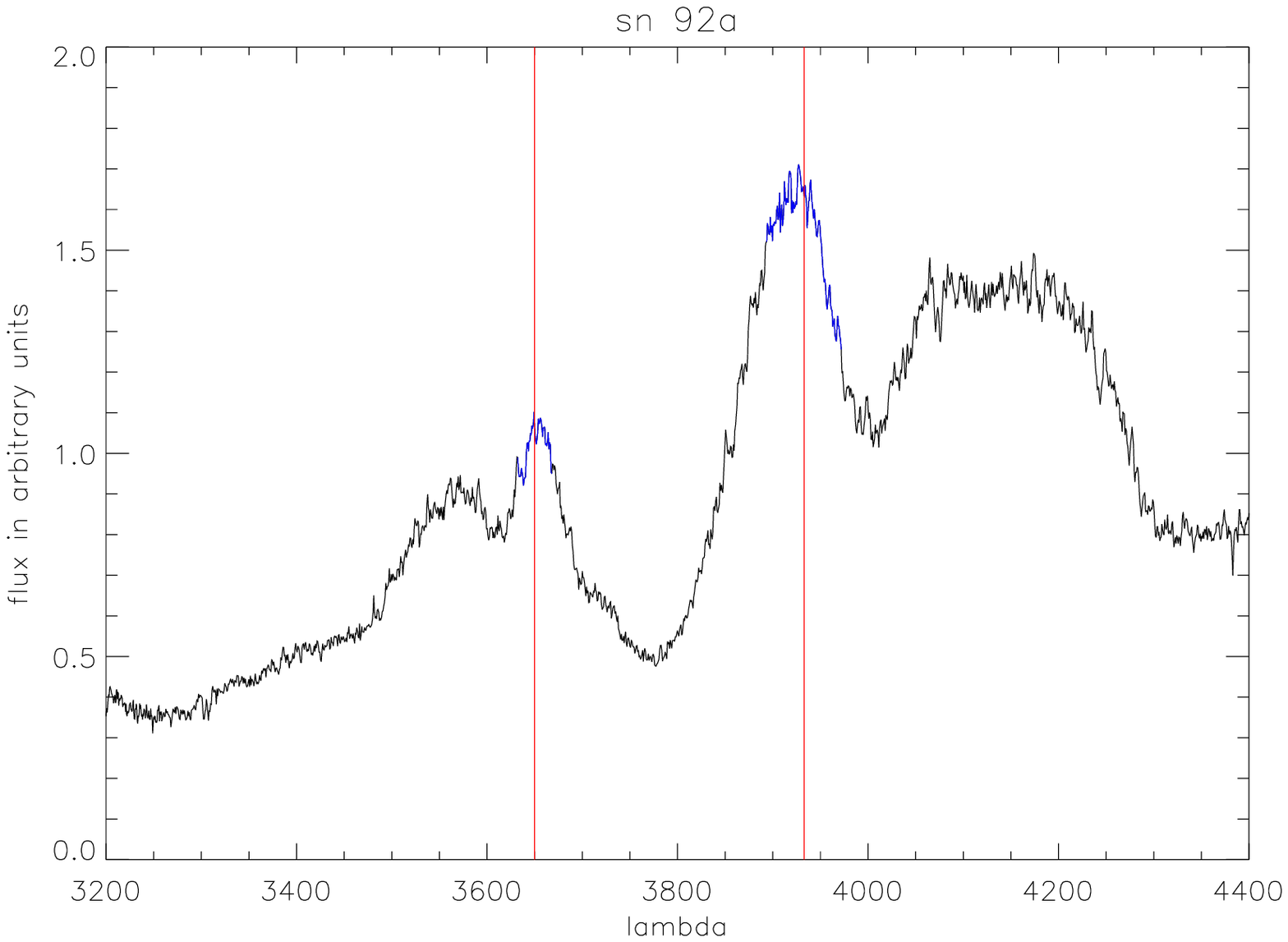}
  \caption[\RCa definition]{The automated construction used to define \RCa. Red
    lines locate  
    the $3650$\AA~ and the $3933$\AA~ \CaII lines. The maxima used to
    calculate \RCa are searched in the two blue colored regions.}
  \label{fig:RCaDef}
\end{figure}

To illustrate the nature that our future research will take, we now
describe some of the past SN~Ia work with \texttt{PHOENIX}.
\citet{nugseq95}, showed that the diversity in the peak of the light
curves of SNe~Ia was correlated with the effective temperature and
likely the nickel mass (see Fig.~\ref{fig:nugseq}). We also showed
that the spectroscopic features of Si~II and Ca~II near maximum light
correlate with the peak brightness of the SN~Ia and that the spectrum
synthesis models by {\tt PHOENIX} were nicely able to reproduce this
effect. We were able to define two spectroscopic indices \RSi and \RCa
(see Figs~\ref{fig:RSiDef}--\ref{fig:RCaDef}), which correlate very
well with the light curve shape parameter \citep{garn99by04}.  These
spectroscopic indices offer an independent (and since they are
intrinsic, they are also reddening independent) approach to
determining peak luminosities of SNe~Ia. S.~Bongard et al.  (in
preparation) have shown that measuring these spectroscopic indicators
may be automated, and that they can be used with the spectral signal
to noise and binning planned for the JDEM missions SNAP and JEDI.

The relationship between the width (and hence risetime) of the
lightcurves of SNe~Ia to the brightness at maximum light is crucial
for precision cosmology. It is well known that the square of the time
difference between 
explosion and peak brightness, $t_{\rm rise}^2$ is proportional to
the opacity, $\kappa$,  \citep{arnett82,byb93}. In an effort to find a
more direct connection between SN~Ia models and the light-curve shape
relationships we examined the Rosseland mean opacity, $\kappa$, at the
center of each model.  We found that in our hotter, more luminous
models $\kappa$ was a factor of 2 times greater than in our cooler,
fainter models. This factor of 1.4 in $t_{\rm rise}$ is very near to
what one would expect, given the available photometric data, for the
ratio of the light-curve shapes between the extremes of SN~1991T (an
over-luminous SN~Ia with a broad light curve) and SN~1991bg (an
under-luminous SN~Ia with a narrow light curve).

We have been studying the effects of evolution on the spectra of
SNe~Ia, in particular the role the initial metallicity of the
progenitor plays in the peak brightness of the SN~Ia. Due to the
effects of metal line blanketing one expects that the metallicity of
the progenitor has a strong influence on the UV spectrum
\citet{hwt98,lentzmet00}.  In \citet{lentzmet00} we quantified these
effects by varying the metallicity in the unburned layers and
computing their resultant spectra at maximum light.

Finally we note the work we have done on testing
detailed hydrodynamical models of SNe~Ia \citep{nughydro97}. 
 It is clear from these calculations that the
sub-Chandrasekhar ``helium-igniter'' models \citep[see for
example][]{wwsubc94} are too 
blue in general and that very few spectroscopic features match the
observed spectrum. On the other hand, the Chandrasekhar-mass model W7 of
\citet{nomw7} is a fairly good match to the early spectra (which are most
important for cosmological applications) of the most typical
SNe~Ia. \citet{l94d01} calculated an extensive time series of W7 and
compared it with that of the well observed nearby SN~Ia SN~1994D. In
this work we showed that W7 fits the observations pretty well at
early times, but the quality of the fits degrades by about 15 days
past maximum light. We speculate that this implies that the outer
layers (seen earliest) of W7 reasonable well represent normal SNe~Ia,
whereas the 
inner layers of SNe~Ia are affected by 3-D mixing effects. With the
work described here, we will be able to directly test this hypothesis
by calculating the spectra of full 3-D hydrodynamical calculations now
being performed by Gamezo and collaborators and by the Munich
group (Hillebrandt and collaborators). \citet{bbbh05}
have calculated very detailed NLTE models of W7
and delayed detonation models of \citet{HGFS99by02}. We find that W7
does not fit the observed Si~II feature very well, although it does a
good job in other parts of the spectrum. The delayed-detonation models
do a bit better, but a highly parameterized model is the best. We
will continue this work as well as extending it to 3-D
models. This will significantly impact our understanding of SNe~Ia 
progenitor, something that is crucial for the success of JDEM.

We stress that the quantitative spectroscopic studies discussed here do
not just show that a proposed explosion model fits or doesn't fit
observed spectra, but provides important information into just how
the spectrum forms. One learns as much from spectra that don't fit as
from ones that do.

Our theoretical work provides important constraints on the science
definition of JDEM, helps to interpret the data coming in now from
both nearby and mid-redshift surveys and involves ongoing code
development to test 3-D hydrodynamic models, as well as both flux and
polarization spectra from nearby supernovae which may indicate
evidence of asphericity. Research such as this requires both manpower and
large-scale computational facilities for production which can be
done to some extent at national facilities such as the National Energy
Research Supercomputing Center at LBNL (NERSC), and local, mid-sized
computing facilities for code development which requires with the
ability to perform tests with immediate turn-around.

%%-----------------------------
%% Bibliography
%%-----------------------------

\clearpage
 
\bibliography{refs,baron,sn1bc,sn1a,sn87a,snii,rte,stars,cosmology,crossrefs}

\end{document}